# Small Teams Propel Fresh Ideas in Science and Technology

**Authors:** Yiling Lin[1,2] and Lingfei Wu[1,2]

**Affiliations:**

[1] School of Computing and Information, The University of Pittsburgh, 135 N Bellefield Ave, Pittsburgh, PA 15213, United States

[2] These authors jointly supervised this work: Yiling Lin, Lingfei Wu. e-mail: yil285@pitt.edu; liw105@pitt.edu

The past half-century has seen a dramatic increase in the scale and complexity of scientific research[1], to which researchers have responded by dedicating more time to education and training[2], narrowing their areas of specialization[3], and collaborating in larger teams[1,3,4]. A widely held view is that such collaborations, by fostering specialization and encouraging novel combinations of ideas, accelerate scientific innovation[5]. However, recent research challenges this notion, suggesting that small teams and solo researchers consistently disrupt science and technology with fresh ideas and opportunities, while larger teams tend to refine existing ones[6]. This study, along with other relevant research[7], has garnered attention for challenging the zeitgeist of our time that views collaboration as the inevitable path forward in scientific and technological advancement[8]. Yet, few studies have re-evaluated its central finding: the innovative advantage of small teams over large ones, using alternative measures.

This study explores innovation by identifying papers proposing new scientific concepts and patents introducing new technology codes. For papers, we use the taxonomy provided by the Microsoft Academic Graph (MAG) team, encompassing a six-level hierarchy[9]. We categorize a paper as introducing a new concept if any taxonomy label, such as "time-evolving block decimation"[10], appears in its title for the first time in our dataset. For patents, we employ the Cooperative Patent Classification (CPC), a four-level system. We categorize a patent as introducing a new technological concept if any taxonomy label, such as "Web crawling techniques for indexing[11], is assigned to it for the first time in the analyzed data.

We analyzed 88 million research articles spanning from 1800 to 2020 and 7 million patent applications from 1976 to 2020 worldwide. Our findings confirm that while large teams contribute to development, small teams play a critical role in innovation by propelling fresh, original ideas in science and technology (see Fig. 1).

This study highlights the innovative advantage of small teams in driving fresh and original ideas in science and technology. Contrary to the prevailing trend emphasizing collaboration as the sole path forward, our findings emphasize the continued importance of smaller, more agile teams in introducing disruptive concepts. Moreover, our research demonstrates the potential for ongoing validation of research findings built upon innovative metrics, such as the Disruption index[6]. Given the ongoing debates about the nature and validity of these metrics[12], it is essential to continue validation using alternative measures of team innovation.

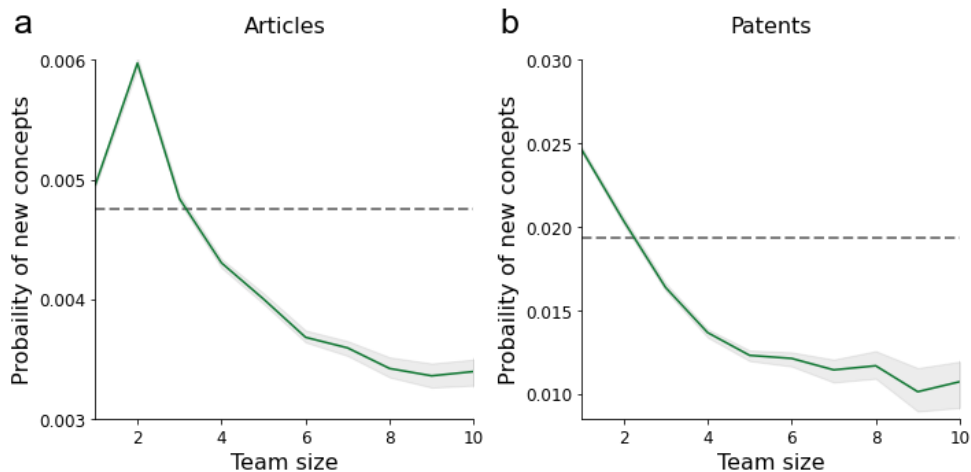

**Figure 1**. Analysis of 87,858,835 scientific papers published between 1800 and 2020 (**a**) and 6,556,163 US patents filed between 1976 and 2020 (**b**) shows that the probability of introducing new concepts within papers and patents decreases with team size. Notably, within papers, this probability increases when the team size grows from one to two and then decreases monotonically. This finding slightly deviates from Wu et al. 2019[6], where the D-index peaks for solo researchers and decreases with additional team members for papers. Nevertheless, the key takeaway from both (a) and (b) is that team size and inventive capability are negatively correlated, in line with the primary findings of Wu et al. 2019[6]. For these two datasets, the Pearson correlation coefficient is -0.002 for papers and -0.03 for patents, with p-values below 0.001 for both. The population probability of introducing new concepts is 0.0047 for papers and 0.019 for patents (grey lines).

## References

1. Wuchty, S., Jones, B. F. & Uzzi, B. The increasing dominance of teams in production of knowledge. *Science* **316**, 1036–1039 (2007).
2. Jones, B. F. The Burden of Knowledge and the 'Death of the Renaissance Man': Is Innovation Getting Harder? *Rev. Econ. Stud.* **76**, 283–317 (2009).
3. Leahey, E. & Reikowsky, R. C. Research Specialization and Collaboration Patterns in Sociology. *Social Studies of Science* vol. 38 425–440 Preprint at https://doi.org/10.1177/0306312707086190 (2008).
4. Jones, B. F. The Rise of Research Teams: Benefits and Costs in Economics. *J. Econ. Perspect.* **35**, 191–216 (2021).
5. Weitzman, M. L. Recombinant growth. *Q. J. Econ.* **113**, 331–360 (1998).
6. Wu, L., Wang, D. & Evans, J. A. Large teams develop and small teams disrupt science and technology. *Nature* **566**, 378–382 (2019).
7. Milojević, S. Quantifying the cognitive extent of science. *J. Informetr.* **9**, 962–973 (2015).
8. Azoulay, P. Small-team science is beautiful. *Nature* **566**, 330–332 (2019).
9. Sinha, A. *et al.* An Overview of Microsoft Academic Service (MAS) and Applications. in *Proceedings of the 24th International Conference on World Wide Web* 243–246 (Association for Computing Machinery, 2015).
10. Orús, R. & Vidal, G. Infinite time-evolving block decimation algorithm beyond unitary evolution. *Phys. Rev. B Condens. Matter* **78**, 155117 (2008).
11. Tamano, M., Okuda, H., Yajima, H., Hirose, T. & Kagaya, N. Retrieval method using image information. *US Patent* (1999).
12. Leibel, C. & Bornmann, L. What do we know about the disruption indicator in scientometrics? An overview of the literature. *arXiv [cs.DL]* (2023).